\documentclass[prd,reprint,showpacs,showkeys]{revtex4-1}
\usepackage{amssymb}
\usepackage{amsmath}
\usepackage{graphicx}
\usepackage[font={footnotesize,it}]{caption}

\begin{document}

\title{Unified Bertotti-Robinson and Melvin Spacetimes}
\author{S. Habib Mazharimousavi}
\email{habib.mazhari@emu.edu.tr}
\author{M. Halilsoy}
\email{mustafa.halilsoy@emu.edu.tr}
\affiliation{Department of Physics, Eastern Mediterranean University, G. Magusa, north
Cyprus, Mersin 10 - Turkey}
\date{\today }

\begin{abstract}
We present a solution for the Einstein-Maxwell (EM) equations which unifies
both the magnetic Bertotti-Robinson (BR) and Melvin (ML) solutions as a
single metric in the axially symmetric coordinates $\left\{ t,\rho
,z,\varphi \right\} $. Depending on the strength of magnetic field the
spacetime manifold, unlike the cases of separate BR and ML spacetime,
develops singularity on the symmetry axis ($\rho =0$). Our analysis shows,
beside other things that there are regions inaccessible to all null
geodesics.
\end{abstract}

\maketitle
\affiliation{Department of Physics, Eastern Mediterranean University, G. Magusa, north
Cyprus, Mersin 10 - Turkey}

\section{Introduction}

The Bertotti-Robinson (BR) \cite{1} and Melvin (ML) \cite{2} solutions of
Einstein-Maxwell (EM) theory are well-known for a long time which had
significant impacts on different aspects of general relativity. For decades
they remained in fashion and found applications in connection with stellar
objects, cosmology, string theory etc. A recent study discusses the
similarities / differences between these spacetimes \cite{3}. It is shown,
among other things in \cite{3} for instance, that the only geodesically
complete static EM spacetimes are the BR and ML solutions. Since they share
\ more common properties than contrasts, the natural question arises whether
it is possible to describe both solutions in a common metric. This is
precisely what we show in the axially symmetric (i.e. $t,$ $\rho ,$ $z,$ $%
\varphi $ coordinates) geometry in this paper. It should be added that large
classes of EM solutions found long ago by Kundt \cite{4} and Plebanski and
Demiahski (PD) \cite{5,6} both admitted separate BR and ML limits in
different coordinates through specific limits. We work out our solution
entirely in the axially symmetric $\left\{ t,\rho ,z,\varphi \right\} $
coordinates and express our metric in those coordinates. Our solution admits
the BR limit but not the separate ML limit. In other words BR universe forms
the background of our spacetime on which ML is added. In obtaining the
solution we choose the magnetic phase of BR solution so that the total
magnetic potential $\psi \left( \rho ,z\right) $ is expressed as a
superposition, $\psi \left( \rho ,z\right) =$ $\psi _{BR}\left( \rho
,z\right) +$ $\psi _{ML}\left( \rho ,z\right) .$ The EM solution constructed
from $\psi \left( \rho ,z\right) $ is what we dub as the "unified BR and ML
spacetime". The solution involves two parameters, $\lambda _{0}$ (for BR
charge) and $B_{0}$ (for ML charge). The ranges of parameters are $%
0<\left\vert \lambda _{0}\right\vert <\infty $ and $-\infty <B_{0}<\infty ,$
so that our solution doesn't admit the ML limit.

BR spacetime is conformally flat whereas ML is cylindrically symmetric which
becomes flat near the axis $\rho \rightarrow 0.$ For a finite $\rho $ and $%
\left\vert z\right\vert \rightarrow \infty $ ML is not flat. Both are
singularity free; a feature that makes them attractive in cosmology and
string theory. We remark also that the BR solution can be obtained by a
coordinate transformation \cite{7} from a spacetime of colliding
electromagnetic waves known as Bell-Szekeres solution \cite{8}. Using the
Ernst formalism we showed long ago that within this formalism Bell-Szekeres
and Khan-Penrose \cite{9} solutions can be combined through a suitable seed
function \cite{10}. Also Schwarzschild and BR spacetimes were interpolated
by the electromagnetic parameter in the oblate spheroidal coordinates \cite%
{11}. Within similar context superposition of spinning spheroids \cite{12}
from harmonic seed functions in the Zipoy-Voorhees metric \cite{13} were
obtained. It is remarkable that interpolation of BR and ML solutions takes
place in the static axial coordinates $\left( \rho ,z\right) $ instead of
oblate/ prolate coordinates. The latter coordinate systems are known to
admit separability in the Laplace equation and had much impact in the
development of solution generation techniques. One of the important
conclusions to be drawn in this study is that two electromagnetic fields,
which separately yield regular spacetimes, namely the BR and ML, may yield a
singular spacetime upon their combination. Physical interpretation suggests
that the mutual magnetic fields focus each other strong enough to result in
a singularity. The singularity at $\rho =z=0$ (for $\frac{B_{0}}{\lambda _{0}%
}<0$ and $\frac{B_{0}}{\lambda _{0}}>1$) doesn't exhibit directional
properties \cite{14}, that is, the Kretschmann scalar diverges irrespective
of the way of approach and ($\rho =0,z>0$) is the only singularity in our
solution for arbitrary parameters. An exact solution of null geodesics
reveals that we have a null-geodesically incomplete manifold. Beside null
geodesics we study the radial motion for massless / massive particles and
also the circular motion in the $z=0$ plane. From the analysis of the
potential the circular motion admits stable orbits.

Organization of the paper is as follows. In Sec. II we introduce magnetic
fields in axial symmetry, solve the equations and derive the metric of
Unified BR and ML spacetimes. Geodesic equation and its solutions are
investigated in Sec. III. The paper ends with Conclusion in Sec. IV.

\section{Magnetic fields in static axial symmetry}

To review the basics of an axially symmetric spacetime we start with the
line element 
\begin{equation}
ds^{2}=-e^{2U}dt^{2}+e^{-2U}\left[ e^{2K}\left( d\rho ^{2}+dz^{2}\right)
+\lambda ^{2}d\varphi ^{2}\right]
\end{equation}%
in which $U,$ $K$ and $\lambda $ are functions of $\rho $ and $z$ alone. The
EM field equations can be derived from a variational principle of the action%
\begin{equation}
I=\int Ld\rho dz
\end{equation}%
where 
\begin{equation}
L=K_{\rho }\lambda _{\rho }+K_{z}\lambda _{z}-\lambda \left[ U_{\rho
}^{2}+U_{z}^{2}-e^{-2U}\left( \psi _{\rho }^{2}+\psi _{z}^{2}\right) \right]
.
\end{equation}%
Here $f_{\rho }$ $/$ $f_{z}$ denotes partial derivative of a function $%
f\left( \rho ,z\right) $ with respect to $\rho $ $/$ $z$ and $\psi $ is a
magnetic potential. Upon variation the metric function $\lambda $ is fixed
as $\lambda =\rho $, while the two basic equations take the forms 
\begin{equation}
\left( \rho U_{\rho }\right) _{\rho }+\rho U_{zz}-\rho e^{-2U}\left( \psi
_{\rho }^{2}+\psi _{z}^{2}\right) =0
\end{equation}%
\begin{equation}
\left( \rho e^{-2U}\psi _{\rho }\right) _{\rho }+\rho \left( e^{-2U}\psi
_{z}\right) _{z}=0.
\end{equation}%
The $K$ function is determined more appropriately by the set%
\begin{equation}
K_{\rho }=\rho \left( U_{\rho }^{2}-U_{z}^{2}\right) +\rho e^{-2U}\left(
\psi _{z}^{2}-\psi _{\rho }^{2}\right)
\end{equation}%
\begin{equation}
K_{z}=2\rho U_{\rho }U_{z}-2\rho e^{-2U}\psi _{\rho }\psi _{z}
\end{equation}%
whose integrability condition is satisfied by virtue of the field equations.
The magnetic vector potential is chosen simply by%
\begin{equation}
A_{\mu }=\delta _{\mu }^{\varphi }\Phi
\end{equation}%
for a function $\Phi \left( \rho ,z\right) $ which is related to $\psi $
above through 
\begin{equation}
\Phi _{\rho }=\rho e^{-2U}\psi _{z}
\end{equation}%
\begin{equation}
\Phi _{z}=-\rho e^{-2U}\psi _{\rho }.
\end{equation}%
The dual of the field tensor $^{\star }F_{ti}=\psi _{i}$ implies the absence
of any electric components which is our choice here. In \cite{3} BR and ML
solutions are summarized in details so that we can only record them in what
follows:

\subsection{BR and ML solutions}

\subsubsection{The BR solution}

\begin{equation}
U=U_{BR}=\ln \lambda _{0}+\frac{1}{2}\ln \left( \rho ^{2}+z^{2}\right)
\end{equation}%
\begin{equation}
\psi =\psi _{BR}=\lambda _{0}\sqrt{\rho ^{2}+z^{2}}
\end{equation}%
\begin{eqnarray}
K &=&K_{BR}=\text{const.} \\
&&\left( \lambda _{0}=\text{constant.}\right)  \notag
\end{eqnarray}%
Note that the more familiar $AdS_{2}\times S^{2}$ version of BR spacetime is
given upon the transformation%
\begin{equation}
\rho =\frac{\sin \theta }{r},
\end{equation}%
\begin{equation}
z=\frac{\cos \theta }{r},
\end{equation}%
by 
\begin{equation}
ds^{2}=\frac{1}{r^{2}}\left( -dt^{2}+dr^{2}\right) +d\theta ^{2}+\sin
^{2}\theta d\varphi ^{2}.
\end{equation}

\subsubsection{The ML solution}

\begin{equation}
U=U_{ML}=\ln \left( 1+\frac{B_{0}^{2}}{4}\rho ^{2}\right)
\end{equation}%
\begin{equation}
\psi =\psi _{ML}=B_{0}z
\end{equation}%
\begin{eqnarray}
K &=&K_{ML}=2\ln \left( 1+\frac{B_{0}^{2}}{4}\rho ^{2}\right) \\
&&\left( B_{0}=\text{constant.}\right)  \notag
\end{eqnarray}

\subsection{A combined BR and ML solution}

We proceed now to combine the foregoing solutions. For this purpose we take
the magnetic potential as the superposition of the two foregoing, namely 
\begin{equation}
\psi =\psi _{BR}+\psi _{ML}=\lambda _{0}\sqrt{\rho ^{2}+z^{2}}+B_{0}z
\end{equation}%
where $B_{0}$ and $\lambda _{0}$ are the constants of ML and BR solutions
which are restricted by $0<\left\vert \lambda _{0}\right\vert <\infty $ and $%
-\infty <B_{0}<\infty $. To get an idea about this superposition we resort
to the axial gauge $A_{\mu }=\left( 0,0,0,A_{\varphi }\right) $\ in flat
space%
\begin{equation}
ds^{2}=-dt^{2}+d\rho ^{2}+dz^{2}+\rho ^{2}d\varphi ^{2}.
\end{equation}%
Let $A_{\varphi }^{\left( 1\right) }=\lambda _{0}\sqrt{\rho ^{2}+z^{2}}$ and 
$A_{\varphi }^{\left( 2\right) }=B_{0}z$ be two magnetic potentials where
both solve the Maxwell equations $\partial _{\mu }F^{\left( i\right) \mu \nu
}=0,$ ($i=1,2$) with $F_{\mu \nu }^{\left( i\right) }=\partial _{\mu }A_{\nu
}^{\left( i\right) }-\partial _{\nu }A_{\mu }^{\left( i\right) }.$ It can be
checked easily that their superposition $A_{\mu }=\left( 0,0,0,A_{\varphi
}^{\left( 1\right) }+A_{\varphi }^{\left( 2\right) }\right) $ solves the
superposed Maxwell equation $\partial _{\mu }\left[ \rho \left( F^{1\mu \nu
}+F^{2\mu \nu }\right) \right] =0.$ Upon this observation we seek an
analogous behavior in the curved spacetime and we find out that indeed it
works with some difference. Integration of the field equations from Eq. (4)
to Eq. (7) yields the following results%
\begin{equation}
e^{U}=F
\end{equation}%
\begin{equation}
e^{K}=\frac{F^{2}}{\rho ^{2}+z^{2}}\left( \frac{\rho ^{1+\frac{B_{0}}{%
2\lambda _{0}}}}{z+\sqrt{\rho ^{2}+z^{2}}}\right) ^{\frac{2B_{0}}{\lambda
_{0}}}
\end{equation}%
where the function $F$ is given by%
\begin{equation}
F=\lambda _{0}\left[ \sqrt{\rho ^{2}+z^{2}}\cosh \left( \frac{B_{0}}{\lambda
_{0}}\ln \rho \right) -z\sinh \left( \frac{B_{0}}{\lambda _{0}}\ln \rho
\right) \right] .
\end{equation}%
It is observed easily that setting $B_{0}=0$ recovers the BR solution with a
charge $\lambda _{0}$. However, the limit $\lambda _{0}=0$ does not exist,
which means that although in flat spacetime our electromagnetic field is a
superposition of BR and ML potentials, in curved spacetime the solution
gives only the BR limit correctly. This is in contrast with the $7-$%
parametric PD class of EM solutions \cite{5,6} which admits electromagnetic
fields even in the flat space limit. In our case existence of the BR is
essential while ML limit can't be interpolated. Let us add also that the
metric functions of PD are expressed in its most generality in quartic
polynomial forms whereas our solution involves decimal powers as well. These
distinctive properties suggest that our solution is not included in the
general class of PD. The two are expressed in different coordinates /
symmetries so that transition between the two for arbitrary cases can't be
expressed in closed forms. More specifically, the type-D metric of PD class
that yields separately the ML and BR limits are as follows: i)%
\begin{multline}
ds_{ML}^{2}=p^{2}\left( -Q\left( p\right) d\bar{t}^{2}+\frac{dq^{2}}{Q\left(
q\right) }\right) + \\
\frac{P\left( p\right) }{p^{2}}d\bar{\sigma}^{2}+\frac{p^{2}}{P\left(
p\right) }dp^{2}.
\end{multline}%
Letting $Q\left( q\right) =1,$ $P\left( p\right) =\frac{4}{B_{0}^{2}}\left(
p-1\right) ,$ ($B_{0}=$constant), $p=1+\frac{B_{0}^{2}}{4}\rho ^{2},$ $q=%
\frac{B_{0}^{2}}{2}z,$ $\bar{\sigma}=\varphi ,$ and an overall scaling gives
the ML metric. ii)%
\begin{multline}
ds_{BR}^{2}=b^{2}\left( -Q\left( p\right) dt^{2}+\frac{dq^{2}}{Q\left(
q\right) }\right) + \\
\gamma ^{2}\left( \frac{P\left( p\right) }{p^{2}}d\sigma ^{2}+\frac{p^{2}}{%
P\left( p\right) }dp^{2}\right) .
\end{multline}%
Letting $b=\gamma =1,$ $Q\left( q\right) =q^{2}=\rho ^{2}+z^{2},$ $P\left(
p\right) =1-p^{2}=\frac{\rho ^{2}}{\rho ^{2}+z^{2}},$ $\sigma =\varphi ,$
gives the BR metric in axial symmetry with a unit charge. It remains to be
seen, however that (25) and (26) follow from the PD class of solutions in
the same coordinate patch i.e. without further transformations in the ($p,q$%
) coordinates.

Furthermore it is worthful to look at the form of invariants of the
spacetime. The complete form of the Kretschmann scalar is complicated enough
that we only give it in a series form around $z=0$ i.e., 
\begin{widetext}
\begin{multline}
\mathcal{K}= \frac{\rho ^{-4\beta ^{2}+4\beta }\left[ A_{1}+A_{2}\rho ^{2\beta
}+A_{3}\rho ^{4\beta }+A_{4}\rho ^{6\beta }+A_{5}\rho ^{8\beta }\right] }{%
\lambda _{0}^{4}\left( 1+\rho ^{2\beta }\right) ^{8}}+ \\
\frac{\rho ^{-4\beta ^{2}+4\beta -1}\left[ B_{1}+B_{2}\rho ^{2\beta
}+B_{3}\rho ^{4\beta }+B_{4}\rho ^{6\beta }+B_{5}\rho ^{8\beta }+B_{6}\rho
^{10\beta }+B_{7}\rho ^{12\beta }+B_{8}\rho ^{14\beta }\right] }{\lambda
_{0}^{4}\left( 1+\rho ^{2\beta }\right) ^{11}}z+  \\
\frac{\rho ^{-4\beta ^{2}+4\beta -2}\left[ C_{1}+C_{2}\rho ^{2\beta
}+C_{3}\rho ^{4\beta }+C_{4}\rho ^{6\beta }+C_{5}\rho ^{8\beta }+C_{6}\rho
^{10\beta }+C_{7}\rho ^{12\beta }+C_{8}\rho ^{14\beta }+C_{9}\rho ^{16\beta
}+C_{10}\rho ^{18\beta }+C_{11}\rho ^{20\beta }\right] }{\lambda
_{0}^{4}\left( 1+\rho ^{2\beta }\right) ^{14}}z^{2}+  \\
\mathcal{O}\left( z^{3}\right) ,  
\end{multline}%
\end{widetext}in which $\beta =\frac{B_{0}}{\lambda _{0}}\neq 0$ and $A_{i}$%
, $B_{i}$ and $C_{i}$ are all some polynomial functions of $\beta $ only.
Having up to second order explicitly is enough to conclude that the solution
is singular at $\rho =0$ and $z\neq 0$ for all values of $\beta .$ This is
due to the term $C_{1}\frac{\rho ^{-4\beta ^{2}+4\beta -2}}{\lambda
_{0}^{4}\left( 1+\rho ^{2\beta }\right) ^{14}}z^{2}$ and $B_{1}\frac{\rho
^{-4\beta ^{2}+4\beta -1}}{\lambda _{0}^{4}\left( 1+\rho ^{2\beta }\right)
^{11}}z$ which for $z\neq 0$ diverge for all $\beta .$ The coefficients $%
C_{1}$ and $B_{1}$ are given explicitly by%
\begin{multline}
B_{1}=2048\left( \beta -\frac{1}{2}\right) \times  \\
\left( \beta ^{6}-3\beta ^{5}+\frac{31}{4}\beta ^{4}-\frac{21}{2}\beta
^{3}+16\beta ^{2}-\frac{45}{4}\beta +\frac{9}{2}\right) 
\end{multline}%
and%
\begin{multline}
C_{1}=256\left( -692\beta ^{5}+380\beta ^{6}+855\beta ^{2}+63\right. - \\
\left. 128\beta ^{7}+1100\beta ^{4}-351\beta -1196\beta ^{3}+32\beta
^{8}\right) 
\end{multline}%
which can not be both zero. At $z=0$ 
\begin{multline}
\mathcal{K}= \\
\frac{\rho ^{-4\beta ^{2}+4\beta }}{\lambda _{0}^{4}\left( 1+\rho ^{2\beta
}\right) ^{8}}\left[ A_{1}+A_{2}\rho ^{2\beta }+A_{3}\rho ^{4\beta
}+A_{4}\rho ^{6\beta }+A_{5}\rho ^{8\beta }\right] 
\end{multline}%
which for regularity at $\rho =0$ we must have 
\begin{equation}
-4\beta ^{2}+4\beta \geq 0.
\end{equation}%
We add that 
\begin{multline}
A_{1}= \\
768-2304\beta +3328\beta ^{2}-2304\beta ^{3}+1792\beta ^{5}+256\beta ^{6}
\end{multline}%
which has no real roots. The condition (31) implies that for $0\leq \beta
\leq 1$ the origin is a regular point. Having clarified the role of the BR
parameter $\lambda _{0}$, i.e. that $\lambda _{0}\neq 0$, so that in the
rest of our analysis we may set $\lambda _{0}=1$ without loss of generality.
In brief for $z=0$ and $\rho =0$ the solution is regular if $0\leq \frac{%
B_{0}}{\lambda _{0}}\leq 1$ and singular for other values of $\frac{B_{0}}{%
\lambda _{0}}$. Once more we recall that $\beta =\frac{B_{0}}{\lambda _{0}}=0
$ corresponds to the BR limit whose Kretschmann scalar is $\frac{8}{\lambda
_{0}^{4}}$ and the solution is regular everywhere. The Maxwell $2-$form of
our solution is expressed by 
\begin{equation}
F=\left( \Phi _{\rho }d\rho +\Phi _{z}dz\right) \wedge d\varphi 
\end{equation}%
where $\Phi _{\rho }$ and $\Phi _{z}$ are defined by (9) and (10). As a
result we obtain for the Maxwell invariants 
\begin{equation}
I_{1}=\frac{1}{2}F_{\mu \nu }F^{\mu \nu }=\lambda _{0}^{2}e^{-2K}\left( 1+%
\frac{B_{0}^{2}}{\lambda _{0}^{2}}+\frac{2B_{0}z}{\lambda _{0}\sqrt{\rho
^{2}+z^{2}}}\right) ,
\end{equation}%
\begin{equation}
I_{2}=\frac{1}{2}F_{\mu \nu }{}^{\star }F^{\mu \nu }=0
\end{equation}%
in which $K$ was found in (23).

Nevertheless the following transformations 
\begin{eqnarray}
\zeta &=&\rho +iz \\
du &=&dt-\rho e^{-2U}d\varphi  \notag \\
dv &=&-2e^{2U}dt  \notag
\end{eqnarray}%
casts (1) into the Kundt form \cite{4} 
\begin{equation}
ds^{2}=du\left( dv+Hdu\right) +P^{-2}d\zeta d\bar{\zeta}
\end{equation}%
in which $H=e^{2U}$ and $P=e^{\left( K-U\right) }.$

\section{Geodesic Motion in cylindrical coordinates}

The geodesic equations for the metric given in (1) are (without loss of
generality we choose $\lambda _{0}=1$)%
\begin{equation}
\frac{d}{ds}\left( \frac{\partial L}{\partial \dot{x}^{\mu }}\right) -\frac{%
\partial L}{\partial x^{\mu }}=0
\end{equation}%
in which 
\begin{equation}
2L=-e^{2U}\dot{t}^{2}+e^{-2U}\left[ e^{2K}\left( \dot{\rho}^{2}+\dot{z}%
^{2}\right) +\rho ^{2}\dot{\varphi}^{2}\right]
\end{equation}%
and a dot means $\frac{d}{ds}.$ From the $t$ and $\varphi $ equations one
finds%
\begin{equation}
\dot{t}=Ee^{-2U},\text{ \ }\dot{\varphi}=\frac{\ell ^{2}}{\rho ^{2}}\text{\ }%
e^{2U}.
\end{equation}%
Using $L=\varepsilon =-1,+1,0$ for timelike, spacelike and null geodesics
the other two equations are%
\begin{multline}
\frac{d}{ds}\left( e^{-2U}e^{2K}\dot{\rho}\right) =-U_{\rho
}E^{2}e^{-2U}+\left( K_{\rho }-U_{\rho }\right) \times \\
\left[ \varepsilon +E^{2}e^{-2U}-\frac{\ell ^{4}}{\rho ^{2}}\text{\ }e^{2U}%
\right] -\frac{\ell ^{4}}{\rho ^{2}}U_{\rho }\text{\ }e^{2U}+\frac{\ell ^{4}%
}{\rho ^{3}}\text{\ }e^{2U}
\end{multline}%
and%
\begin{multline}
\frac{d}{ds}\left( e^{-2U}e^{2K}\dot{z}\right) =-U_{z}E^{2}e^{-2U}+\left(
-U_{z}+K_{z}\right) \times \\
\left[ \varepsilon +E^{2}e^{-2U}-\frac{\ell ^{4}}{\rho ^{2}}\text{\ }e^{2U}%
\right] -\frac{\ell ^{4}}{\rho ^{2}}U_{z}\text{\ }e^{2U}.
\end{multline}%
We parametrize now $\rho $ with $z$ so that $\rho ^{\prime }=\frac{d\rho }{dz%
}$and express geodesics in a single equation%
\begin{multline}
\left( E^{2}e^{-2U}+\Delta +\frac{\ell ^{4}}{\rho ^{2}}\text{\ }%
e^{2U}\right) \left( U_{\rho }-\rho ^{\prime }U_{z}\right) - \\
\Delta \left( K_{\rho }-K_{z}\rho ^{\prime }\right) -+\frac{\ell ^{4}}{\rho
^{3}}\text{\ }e^{2U}+\frac{\Delta }{1+\rho ^{\prime 2}}\rho ^{\prime \prime
}=0,
\end{multline}%
where $\Delta =\varepsilon +E^{2}e^{-2U}-\frac{\ell ^{4}}{\rho ^{2}}$\ $%
e^{2U}.$ Let's consider the null $\left( \varepsilon =0\right) $ geodesics
in a plane of $\varphi =\varphi _{0}$ which implies $\ell =0$ and therefore
(43) becomes (with $E^{2}=1$)%
\begin{equation}
\left( 2U_{\rho }-K_{\rho }\right) -\rho ^{\prime }\left(
2U_{z}-K_{z}\right) +\frac{\rho ^{\prime \prime }}{1+\rho ^{\prime 2}}=0.
\end{equation}%
The explicit form of latter equation reads as 
\begin{multline}
\frac{d^{2}\rho }{dz^{2}}=2\left( 1+\left( \frac{d\rho }{dz}\right)
^{2}\right) \times \\
\left[ \frac{z}{\rho ^{2}+z^{2}}\left( \frac{d\rho }{dz}-\frac{\rho }{z}%
\right) +\frac{B_{0}}{\sqrt{\rho ^{2}+z^{2}}}\left( \frac{d\rho }{dz}+\frac{z%
}{\rho }\right) +\frac{B_{0}^{2}}{2\rho }\right]
\end{multline}%
which is still complicated enough for an exact solution. Luckily we obtain
an exact solution valid for $\left\vert B_{0}\right\vert >2$ given by 
\begin{equation}
\rho =\left\{ 
\begin{array}{ccc}
\left( \left( \frac{B_{0}}{2}\right) ^{2}-1\right) ^{1/2}z, & B_{0}<-2, & z>0
\\ 
-\left( \left( \frac{B_{0}}{2}\right) ^{2}-1\right) ^{1/2}z, & B_{0}>2, & z<0%
\end{array}%
\right. .
\end{equation}%
As we stated above $\left\vert B_{0}\right\vert >2$ yields singularity at $%
\rho =z=0$, and our particular solution is valid only for this case. For
each given $\left\vert B_{0}\right\vert >2$ we have a wedge region of $%
\left( \rho ,z\right) $ which doesn't cover all the $\left( \rho ,z\right) $
plane. This is the indication that our particular solution doesn't yield a
null-geodesically complete spacetime. The BR and ML spacetimes are known
both to be geodesically complete whereas our particular example provides a
case of their combination which is at least null--geodesically incomplete.
This can be observed by resorting to the solution (46) to obtain (let us
choose $B_{0}=-4,$ for simplicity)%
\begin{equation}
\frac{d\rho }{d\lambda }=\frac{const.}{\rho ^{8}\left( 3\rho ^{8}+1\right)
^{2}}
\end{equation}%
where $s$ is the affine parameter for null geodesics. A similar equation
follows also for $\frac{dz}{d\lambda }$. Eq. (47) yields a highly localized
solution for $\rho \left( \lambda \right) $ (and $z\left( \lambda \right) $)
justifying the expected incompleteness. Another interesting solution for
(45) can be found exactly when $B_{0}=\pm 1.$ The solution in this case is a
circle of arbitrary radius $a$ in the plane of $\left( \rho ,z\right) $ with
equation $\rho ^{2}+z^{2}=a^{2}.$ Fig. 1 displays the numerical plot from
Eq. (45) for specific initial conditions.

\begin{figure}[tbp]
\includegraphics[width=60mm,scale=0.7]{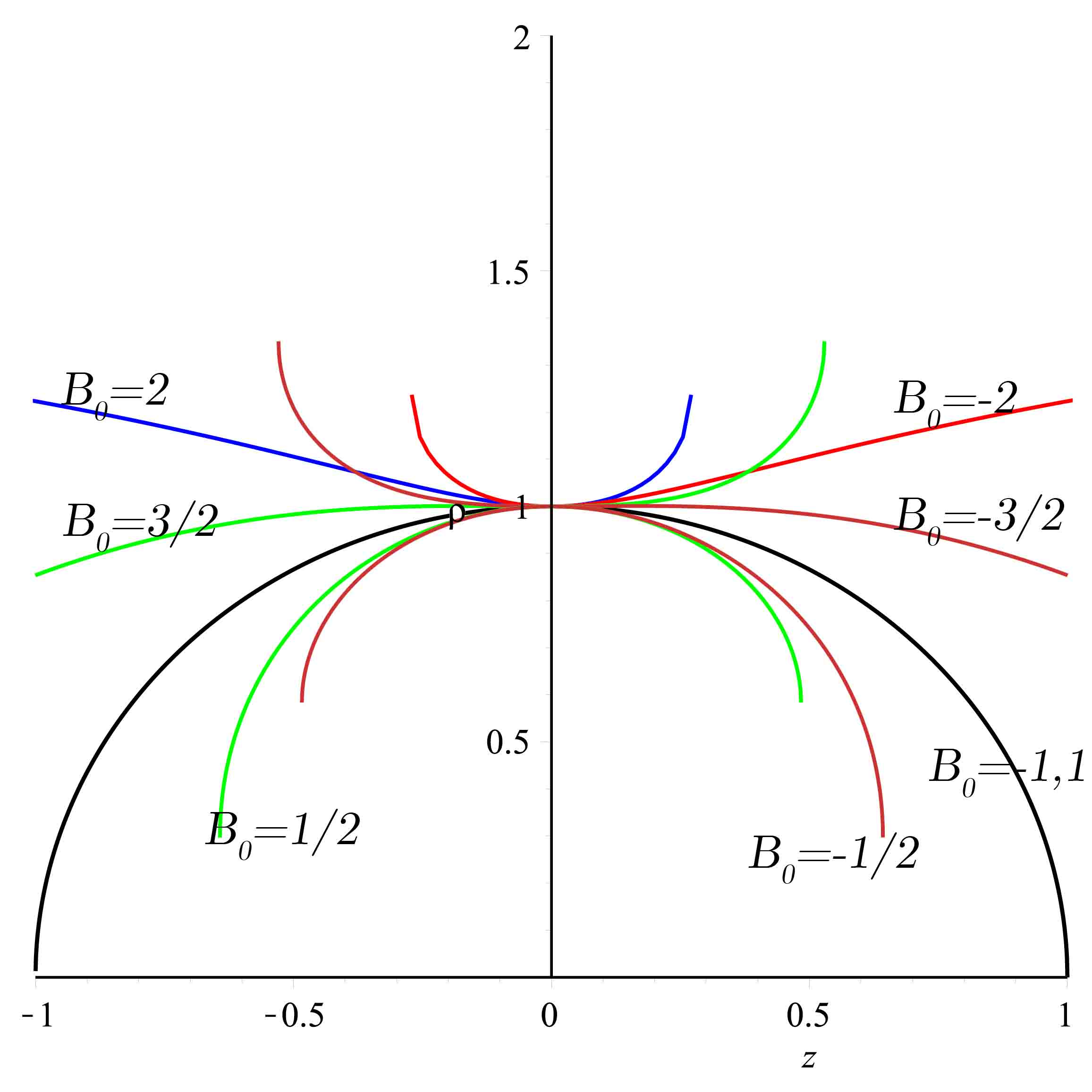}
\caption{Plotting of $\protect\rho \left( z\right) $ versus $z$ in
accordance with the geodesics equation (45), for specific values of $B_{0}$ (%
$\protect\lambda _{0}=1$). The initial conditions are chosen such that $%
\protect\rho \left( 0\right) =1$ and $\protect\rho ^{\prime }\left( 0\right)
=0.$}
\end{figure}

\subsection{Geodesic Motion in $z=0$ plane for $B_{0}=\protect\lambda _{0}$}

As we have shown above the plane $z=0$ has no singularity if $0\leq \frac{%
B_{0}}{\lambda _{0}}\leq 1$. This makes it to be distinguished from the
other planes $z=z_{0}\neq 0.$ Setting $\frac{B_{0}}{\lambda _{0}}=1$ is also
the only value in this interval which makes the power of $\rho $ integer.
Therefore, we are interested to consider the geodesic motion of a massive
particle with unit mass in this spacetime i.e. 
\begin{multline}
ds^{2}=-\frac{\lambda _{0}^{2}\left( \rho ^{2}+1\right) ^{2}}{4}dt^{2}+ \\
\frac{\lambda _{0}^{2}\left( \rho ^{2}+1\right) ^{2}}{4\rho ^{2}}d\rho ^{2}+%
\frac{4}{\lambda _{0}^{2}}\frac{\rho ^{2}}{\left( \rho ^{2}+1\right) ^{2}}%
d\varphi ^{2}.
\end{multline}%
The Lagrangian is given by%
\begin{multline}
\mathcal{L}=-\frac{\lambda _{0}^{2}\left( \rho ^{2}+1\right) ^{2}}{8}\dot{t}%
^{2}+ \\
\frac{\lambda _{0}^{2}\left( \rho ^{2}+1\right) ^{2}}{8\rho ^{2}}\dot{\rho}%
^{2}+\frac{2}{\lambda _{0}^{2}}\frac{\rho ^{2}}{\left( \rho ^{2}+1\right)
^{2}}\dot{\varphi}^{2},
\end{multline}%
in which an over dot shows derivative wrt the affine parameter $\lambda $.
The conservation of energy and angular momentum is obvious such that%
\begin{equation}
\frac{\partial \mathcal{L}}{\partial \dot{t}}=-\frac{\lambda _{0}^{2}\left(
\rho ^{2}+1\right) ^{2}}{4}\dot{t}=-E
\end{equation}%
and 
\begin{equation}
\frac{\partial \mathcal{L}}{\partial \dot{\varphi}}=\frac{4}{\lambda _{0}^{2}%
}\frac{\rho ^{2}}{\left( \rho ^{2}+1\right) ^{2}}\dot{\varphi}=\ell .
\end{equation}%
Having $g_{\mu \nu }\frac{dx^{\mu }}{d\lambda }\frac{dx^{\nu }}{d\lambda }%
=-\epsilon $ where $\epsilon =0$ /$1$ (for unit mass) yields the null or
timelike geodesics, implies%
\begin{equation}
-\frac{\lambda _{0}^{2}\left( \rho ^{2}+1\right) ^{2}}{4}\dot{t}^{2}+\frac{%
\lambda _{0}^{2}\left( \rho ^{2}+1\right) ^{2}}{4\rho ^{2}}\dot{\rho}^{2}+%
\frac{4}{\lambda _{0}^{2}}\frac{\rho ^{2}}{\left( \rho ^{2}+1\right) ^{2}}%
\dot{\varphi}^{2}=-\epsilon
\end{equation}%
or upon using the conserved quantities one finds%
\begin{equation}
\dot{\rho}^{2}=\frac{16\rho ^{2}E^{2}}{\lambda _{0}^{4}\left( \rho
^{2}+1\right) ^{4}}-\frac{4\rho ^{2}\epsilon }{\lambda _{0}^{2}\left( \rho
^{2}+1\right) ^{2}}-\ell ^{2}.
\end{equation}

\subsubsection{Radial motion of massive particle}

Let's consider, as the first case, the motion with zero angular momentum of
a massive particle i.e., $\ell =0$ and $\epsilon =1.$ These in turn yield%
\begin{equation}
\dot{\rho}^{2}=\frac{16\rho ^{2}E^{2}}{\lambda _{0}^{4}\left( \rho
^{2}+1\right) ^{4}}-\frac{4\rho ^{2}}{\lambda _{0}^{2}\left( \rho
^{2}+1\right) ^{2}}
\end{equation}%
which after getting help from 
\begin{equation}
\dot{\rho}^{2}=\left( \frac{\partial \rho }{\partial \lambda }\right)
^{2}=\left( \frac{\partial \rho }{\partial t}\right) ^{2}\left( \frac{%
\partial t}{\partial \lambda }\right) ^{2}=\left( \frac{\partial \rho }{%
\partial t}\right) ^{2}\frac{16E^{2}}{\lambda _{0}^{4}\left( \rho
^{2}+1\right) ^{4}}
\end{equation}%
one finds%
\begin{equation}
\left( \frac{\partial \rho }{\partial t}\right) ^{2}=\rho ^{2}-\frac{\lambda
_{0}^{2}\rho ^{2}\left( \rho ^{2}+1\right) ^{2}}{4E^{2}}.
\end{equation}%
Nevertheless, one may set the affine parameter to be the proper time $\tau $
and therefore%
\begin{equation}
\left( \frac{\partial \rho }{\partial \tau }\right) ^{2}=\frac{16\rho
^{2}E^{2}}{\lambda _{0}^{4}\left( \rho ^{2}+1\right) ^{4}}-\frac{4\rho ^{2}}{%
\lambda _{0}^{2}\left( \rho ^{2}+1\right) ^{2}}.
\end{equation}%
Now suppose the particle starts from rest at $\rho =\rho _{0}$ where $\frac{%
\partial \rho }{\partial t}=$ $\frac{\partial \rho }{\partial \tau }=0,$
which gives%
\begin{equation}
E^{2}=\frac{\lambda _{0}^{2}\left( \rho _{0}^{2}+1\right) ^{2}}{4}.
\end{equation}%
Hence the equations of motion become%
\begin{equation}
\left( \frac{\partial \rho }{\partial t}\right) ^{2}=\rho ^{2}-\frac{\rho
^{2}\left( \rho ^{2}+1\right) ^{2}}{\left( \rho _{0}^{2}+1\right) ^{2}}
\end{equation}%
and%
\begin{equation}
\left( \frac{\partial \rho }{\partial \tau }\right) ^{2}=\frac{4\rho
^{2}\left( \rho _{0}^{2}+1\right) ^{2}}{\lambda _{0}^{2}\left( \rho
^{2}+1\right) ^{4}}-\frac{4\rho ^{2}}{\lambda _{0}^{2}\left( \rho
^{2}+1\right) ^{2}}.
\end{equation}%
In Fig. 2 we plot $\rho $ versus $t$ (a) and $\tau $ (b). It is very much
clear that the motion is periodic which means that the particle is attracted
by the origin and while approaches the origin it gains energy and this
energy causes to pass the origin and in the other direction slows down to
rest and in the same way repeats the motion. The different between the
period of motion measured by an observer on the particle and observer in the
lab is also manifested in the figures.

\begin{figure}[tbp]
\includegraphics[width=60mm,scale=0.7]{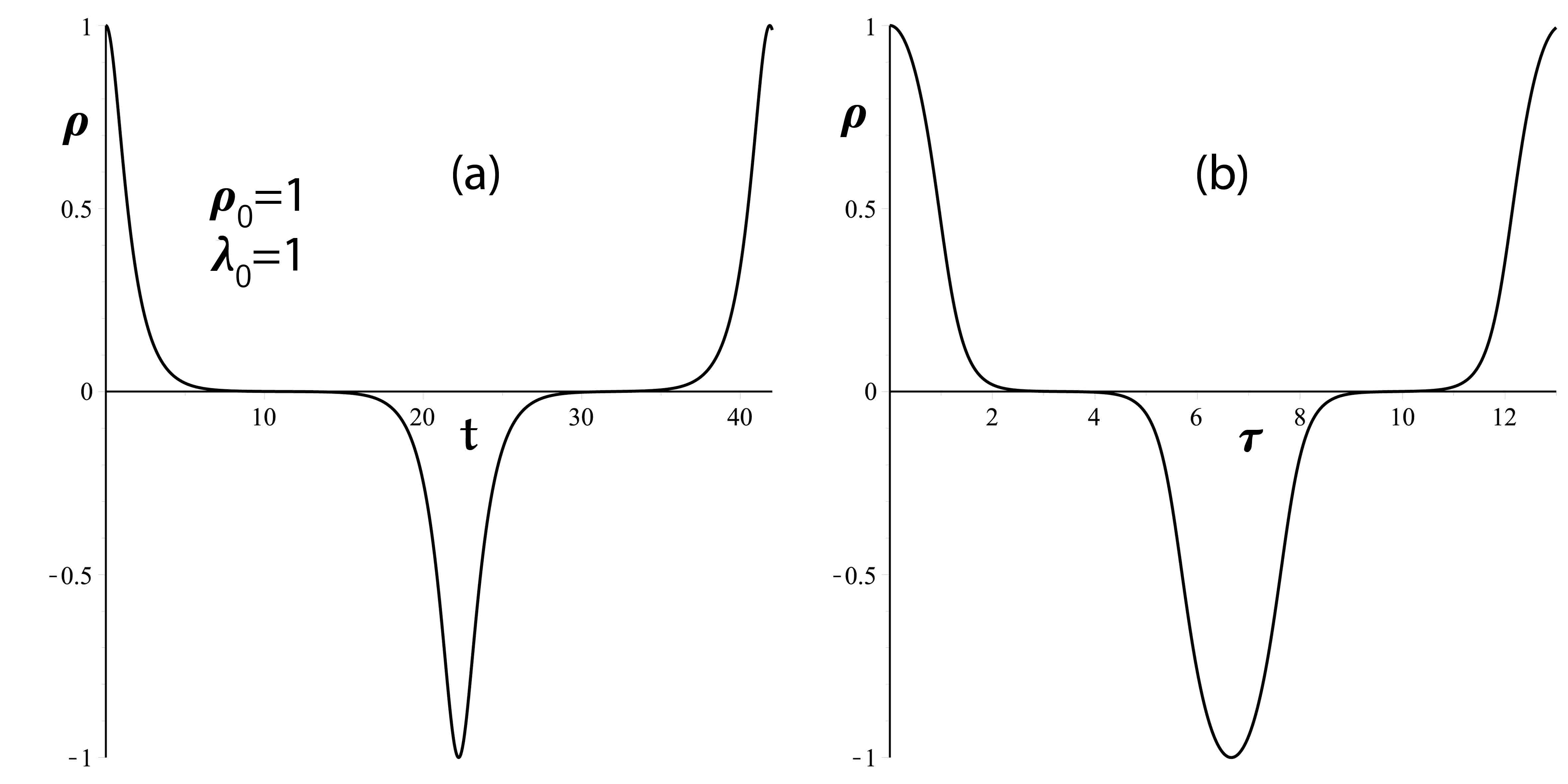}
\caption{Radial fall from $\protect\rho =1$ through $\protect\rho =0$ (in
the $z=0$ plane) for a fixed angle as a function of the coordinate time
(Fig. 2a) / proper time (Fig. 2b). The particle crosses $\protect\rho =0$
freely since $\protect\rho =0=z$ is not singular in the chosen interval $%
\frac{B_{0}}{\protect\lambda _{0}}=1$.}
\end{figure}

\subsubsection{Radial motion of a massless particles}

In the same way one may study the motion of a null particle with $\epsilon
=0 $ and $\ell =0.$ The equation of motion Eq. (53) then reads%
\begin{equation}
\dot{\rho}^{2}=\frac{16\rho ^{2}E^{2}}{\lambda _{0}^{4}\left( \rho
^{2}+1\right) ^{4}}
\end{equation}%
which after using the chain rule we find%
\begin{equation}
\left( \frac{d\rho }{dt}\right) ^{2}=\rho ^{2}.
\end{equation}%
whose explicit solution is given by%
\begin{equation}
\rho =\rho _{0}e^{\pm t}
\end{equation}%
where $\pm $ refers to the direction of motion.

\subsubsection{Circular Motion}

To work out the circular motion of a particle on the plane $z=0$ we use the
chain rule in (53) to find%
\begin{multline}
\left( \frac{d\rho }{d\varphi }\right) ^{2}= \\
\frac{16^{2}\rho ^{6}E^{2}}{\lambda _{0}^{8}\left( \rho ^{2}+1\right)
^{8}\ell ^{2}}-\frac{64\rho ^{6}\epsilon }{\lambda _{0}^{6}\left( \rho
^{2}+1\right) ^{6}\ell ^{2}}-\frac{16\rho ^{4}}{\lambda _{0}^{4}\left( \rho
^{2}+1\right) ^{4}}.
\end{multline}%
As usual we introduce $u=\frac{1}{\rho }$ to change the equation of motion
in the form of%
\begin{multline}
\left( \frac{du}{d\varphi }\right) ^{2}=\frac{256u^{14}E^{2}}{\lambda
_{0}^{8}\left( u^{2}+1\right) ^{8}\ell ^{2}}- \\
\frac{64u^{10}\epsilon }{\lambda _{0}^{6}\left( u^{2}+1\right) ^{6}\ell ^{2}}%
-\frac{16u^{8}}{\lambda _{0}^{4}\left( u^{2}+1\right) ^{4}}=A\left( u\right)
\end{multline}%
Having a photon ($\epsilon =0$) or a massive particle ($\epsilon =1$) moving
on a circular orbit means $\left. A\left( u\right) \right\vert _{u_{c}}=0$
and having an equilibrium path needs an additional condition $\left. \frac{%
dA\left( u\right) }{du}\right\vert _{u_{c}}=0.$ Herein $\rho _{c}=\frac{1}{%
u_{c}}$is the radius of the equilibrium circular orbit. For the massive
particle ($\epsilon =1$) these conditions yield%
\begin{equation}
E^{2}=\frac{\left( u_{c}^{2}-1\right) \left( u_{c}^{2}+1\right) ^{2}\lambda
_{0}^{2}}{4\left( u_{c}^{2}-3\right) u_{c}^{4}}
\end{equation}%
and%
\begin{equation}
\ell ^{2}=\frac{8u_{c}^{2}}{\lambda _{0}^{2}\left( u_{c}^{2}+1\right)
^{2}\left( u_{c}^{2}-3\right) }
\end{equation}%
and therefore the radius of the circular path is found to be the positive
root of the following equation%
\begin{equation}
\left( u_{c}^{4}-1\right) \left( u_{c}^{2}+1\right) ^{4}=\frac{32E^{2}}{%
\lambda _{0}^{4}\ell ^{2}}u_{c}^{6}.
\end{equation}%
From the latter equation we see that for $E=0$ a circular path with $u_{c}=1$
is possible. This is in fact the maximum value of the possible radius for a
circular motion. Particles with higher energy may be able to orbit about the
origin with a radius less then one.

For a massless particle the same conditions dictate a single circular path
with 
\begin{equation}
u_{c}=\sqrt{3}
\end{equation}%
and energy satisfying 
\begin{equation}
\frac{E^{2}}{\ell ^{2}}=\frac{16}{27}\lambda _{0}^{4}.
\end{equation}

\begin{figure}[tbp]
\includegraphics[width=60mm,scale=0.7]{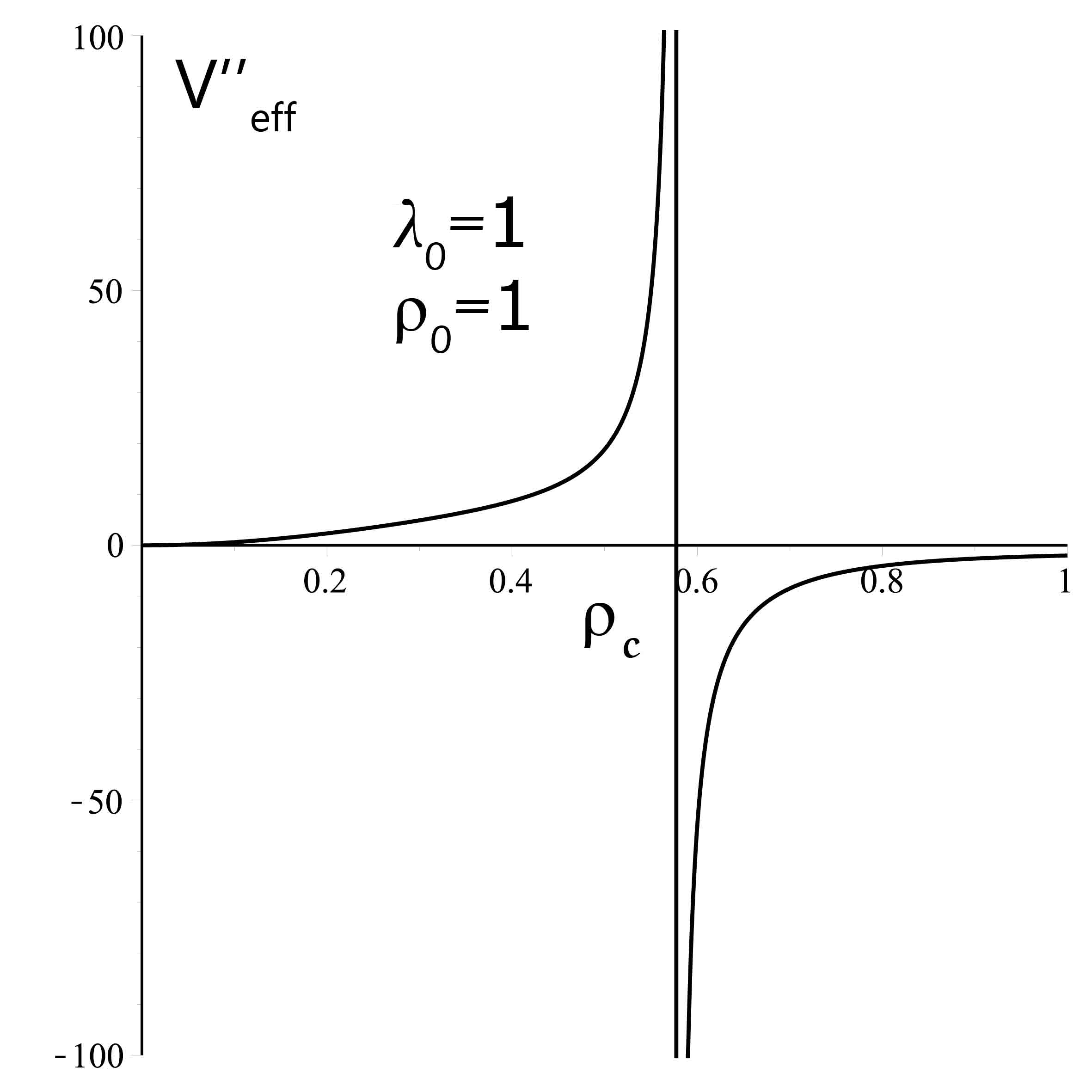}
\caption{Stability condition for particles in circular orbits. From Eq.
(74), $V^{\prime \prime }\left( \protect\rho _{c}\right) $ is plotted versus
the circular radius $\protect\rho _{c}.$ It is observed that for $\protect%
\rho _{c}<\frac{1}{\protect\sqrt{3}}$ we have stable orbits since it gives $%
V^{\prime \prime }\left( \protect\rho _{c}\right) >0$. }
\end{figure}

\subsubsection{Stability of the circular motion}

To see whether the circular path of the particles found above are stable or
not we go back to the Eq. (54) and rewrite it in the form of one dimensional
motion%
\begin{gather}
\frac{1}{2}\left( \frac{d\rho }{d\lambda }\right) ^{2}+V_{eff}=0, \\
V_{eff}=-\text{ }\frac{8\rho ^{2}E^{2}}{\lambda _{0}^{4}\left( \rho
^{2}+1\right) ^{4}}+\frac{2\rho ^{2}\epsilon }{\lambda _{0}^{2}\left( \rho
^{2}+1\right) ^{2}}+\frac{\ell ^{2}}{2}.
\end{gather}%
An expansion of $V_{eff}$ about $\rho =\rho _{c}$ yields (we note that at
equilibrium circular path both $V_{eff}$ and its first derivative vanish)%
\begin{equation}
\left( \frac{dx}{d\lambda }\right) ^{2}+V_{eff}^{\prime \prime }\left( \rho
_{c}\right) x^{2}=0
\end{equation}%
where $x=\rho -\rho _{c}$ and 
\begin{equation}
V_{eff}^{\prime \prime }\left( \rho _{c}\right) =\frac{16\left(
2u_{c}^{4}-3u_{c}^{2}+3\right) u_{c}^{4}}{\left( u_{c}^{2}-3\right) \left(
u_{c}^{2}+1\right) ^{4}\lambda _{0}^{2}}.
\end{equation}%
A second derivative wrt $\lambda $ from (71) admits%
\begin{equation}
\left( \frac{d^{2}x}{d\lambda ^{2}}\right) +V_{eff}^{\prime \prime }\left(
\rho _{c}\right) x=0
\end{equation}%
which has an oscillatory motion of $x$ wrt $\lambda $ (stable motion) if $%
V_{eff}^{\prime \prime }\left( \rho _{c}\right) >0.$ Fig. 3 displays $%
V_{eff}^{\prime \prime }\left( \rho _{c}\right) $ versus $\rho _{c}.$ As it
is clear those orbits whose radius is less then $\frac{1}{\sqrt{3}}$ are
stable. Similar argument can be repeated for the massless particles. The
effective potential and its first derivative at $\rho =\rho _{c}=\frac{1}{%
\sqrt{3}}$ are zero while%
\begin{equation}
V_{eff}^{\prime \prime }\left( \rho _{c}\right) =\frac{243E^{2}}{32\lambda
_{0}^{4}}
\end{equation}%
which is clearly positive. Therefore the orbit of a photon is stable which
is unlike the Schwarzschild and Reissner-Nordstr\"{o}m spacetime.

\subsubsection{Null Geodesics in Kundt form}

The Lagrangian of an uncharged particle moving in the spacetime identified
by (37) reads as%
\begin{equation}
L=\dot{u}\dot{v}+H\dot{u}^{2}+e^{2\left( K-U\right) }\left( \dot{\rho}^{2}+%
\dot{z}^{2}\right)
\end{equation}%
in which ($\cdot \equiv \frac{d}{d\lambda }$). The first equation $\frac{d}{%
d\lambda }\left( \frac{\partial L}{\partial \dot{v}}\right) =\frac{\partial L%
}{\partial v}$ yields 
\begin{equation}
\ddot{u}=0
\end{equation}%
which in turn implies $\dot{u}=$constant. This basically suggests that our
affine parameter $\lambda $ is $u.$ The second equation $\frac{d}{d\lambda }%
\left( \frac{\partial L}{\partial \dot{u}}\right) =\frac{\partial L}{%
\partial u}$ gives%
\begin{equation}
\frac{dv}{du}+2H=\alpha _{0}
\end{equation}%
where $\alpha _{0}$ is an integration constant. The other two equations are
also given by%
\begin{equation}
\ddot{\rho}+\left( K-U\right) _{\rho }\left( \dot{\rho}^{2}-\dot{z}%
^{2}\right) +2\dot{\rho}\dot{z}\left( K-U\right) _{z}=\dot{\rho}U_{\rho
}e^{2\left( 2U-K\right) }
\end{equation}%
and%
\begin{equation}
\ddot{z}+\left( K-U\right) _{z}\left( \dot{z}^{2}-\dot{\rho}^{2}\right) +2%
\dot{\rho}\dot{z}\left( K-U\right) _{\rho }=\dot{z}U_{z}e^{2\left(
2U-K\right) }
\end{equation}%
in which herein ($\cdot \equiv \frac{d}{du}$). \ For $\varphi =$constant one
finds $du=dt,$ and the equation (79) is satisfied if $\alpha _{0}=0$. For
null-geodesics we find from (77) that%
\begin{equation}
\left( \rho _{u}^{2}+z_{u}^{2}\right) =e^{2\left( 2U-K\right) }
\end{equation}%
and upon the symmetry between $\rho $ and $z$ we set $\rho =\kappa z$ with $%
\kappa =$constant to get (we choose also $\lambda _{0}=1$)%
\begin{equation}
\frac{dz}{e^{2U-K}}=\frac{du}{\sqrt{1+\kappa ^{2}}}
\end{equation}%
in which $du=dt.$ A substitution and integration admits%
\begin{eqnarray}
z &=&\left( \frac{B_{0}^{2}-1}{k_{0}}\left( t-t_{0}\right) \right) ^{\frac{1%
}{B_{0}^{2}-1}} \\
&&\left( B_{0}^{2}\neq 1\right)  \notag
\end{eqnarray}%
where $k_{0}=\frac{\kappa ^{2B_{0}+B_{0}^{2}}}{\sqrt{1+\kappa ^{2}}\left( 1+%
\sqrt{1+\kappa ^{2}}\right) ^{2B_{0}}}$ and $t_{0}$ is an integration
constant. For $B_{0}^{2}=1$ we find%
\begin{equation}
k_{0}\ln z=t-t_{0}.
\end{equation}%
This brief analysis of Kundt's null geodesics recovers the equivalent
results of the previous analysis. Namely, that the exact integrals of
geodesics in a section of the ($\rho ,z$) plane doesn't cover the whole
plane. We conclude therefore that null geodesic incompleteness remains
intact irrespective of the representation of the metric.

\section{Conclusion}

Being inspired by the superposed solutions in colliding wave spacetimes
which unfortunately received no attentions we show here in a similar manner
that BR and ML spacetimes can be combined in a single metric. The
distinction between the two problems, i.e. colliding waves and axial
symmetry, is that in the latter case superposition worked in the more
familiar cylindrical $\left( \rho ,z\right) $ coordinates rather than the
prolate / oblate ones. The obtained metric inherits the imprints of both
solutions. It is not conformally flat for instance, and regularity at the
origin i.e. at $\rho =z=0,$ holds provided in $0\leq \frac{B_{0}}{\gamma _{0}%
}\leq 1.$ For an arbitrary ML parameter, however, our solution becomes
singular on the symmetry axis. Due to the fractional powers of $\rho $ our
solution is neither smooth nor flat on the symmetry axis. Exact solution of
geodesics reveals that null geodesics in the singular manifold are not
complete whereas BR and ML spacetimes separately are known to admit complete
geodesics. One drawback of our solution is that $\gamma _{0}\rightarrow 0$
limit i.e. the ML limit doesn't exist. In a single coordinate patch the
large type-D Einstein-Maxwell family of Plebanski and Demianski (PD) also
suffers a similar problem. In this regard our overall impression is that our
non-smooth solution doesn't belong to the class of PD. Finally we add that
this simple example may serve to pave the way for further 'superposed'
spacetimes in general relativity, including the higher dimensional ones.

\begin{acknowledgments}
We wish to thank the anonymous referee for much valuable suggestions.
\end{acknowledgments}


\begin{thebibliography}{99}
\bibitem{1} B. Bertotti, Phys. Rev. \textbf{116}, 1331 (1959);

I. Robinson, Bull. Acad. Pol. Sci. Ser. Sci. Math. Astr. Phys. \textbf{7},
351 (1959).

\bibitem{2} M. A. Melvin, Phys. Letters \textbf{8}, 65 (1964).

W. B. Bonnor, Proc. Phys. Soc. A \textbf{67}, 225 (1954).

\bibitem{3} D. Garfinkle and E. N. Glass, Class. Quantum Grav. \textbf{28},
215012 (2011).

\bibitem{4} W. Kundt, Proc. Roy. Soc. A \textbf{270}, 328 (1962).

\bibitem{5} J. F. Plebanski and M. Demiahski, Rotating, charged, and
uniformly accelerating mass in general relativity, Ann. Phys. 98, 98 (1976);

J. Pleba\'{n}ski, J. Math. Phys. 20, 1946 (1979);

J. B. Griffiths and J. Podolsk\'{y}, Int. J. Mod. Phys. D \textbf{15}, 335
(2006).

\bibitem{6} J. B. Griffiths and J. Podolsk\'{y}, "Exact Space-Times in
Einstein's General Relativity", Cambridge Monographs on Mathematical
Physics, Cambridge University Press, Cambridge U.K. (2009).

\bibitem{7} S. Chandrasekhar and B. C. Xanthopoulos Proc. Roy. Soc. A 
\textbf{140}, 311 (1987).

\bibitem{8} P. Bell and P. Szekeres, Gen. Rel. Grav. \textbf{5}, 275 (1974).

\bibitem{9} K. A. Khan and R. Penrose, Nature, \textbf{229}, 185 (1971).

\bibitem{10} M. Halilsoy, J. Math. Phys. \textbf{34}, 3553 (1993).

\bibitem{11} M. Halilsoy, Gen. Rel. Grav. \textbf{25}, 275 (1992).

\bibitem{12} M. Halilsoy, J. Math. Phys. \textbf{33}, 4225 (1992).

\bibitem{13} D. M. Zipoy, J. Math. Phys. \textbf{7}, 1137 (1966)

B. H. Voorhees, Phys. Rev. D. \textbf{2}, 2119 (1970).

\bibitem{14} R. Gautreau and J. L. Anderson, Phys. Lett. A \textbf{25}, 291
(967).
\end{thebibliography}
\end{document}